\documentstyle[12pt]{article}
\begin{document}

\begin{large}
\begin{center}

{\bf Phase diagram of the pairing symmetry
     in two-dimensional strong-coupling superconductors}

\end{center}
\end{large}

\vspace{3mm}

\begin{small}
\begin{center}
{Katsuji Sakai, Yasushi Yokoya and Yoshiko Oi Nakamura}
\end{center}
\begin{center}
{\it Department of Applied Physics, Faculty of Science,
 Science University of Tokyo,\\
 1-3 Kagurazaka, Shinjuku-ku, Tokyo 162, Japan}
\end{center}

\end{small}


\begin{abstract}

Two-dimensional Eliashberg equations
 have been solved
 by use of a mixed interaction with $s$- and $d$-channels.
It is discussed
 what kind of pairing symmetry of the superconducting state
 can be realized
 when the channel mixing parameters and the band-filling
 are varied.
By changing the mixing parameters
 and varying the chemical potential $\mu$
 between zero- and half-filling,
 the pairing symmetry is determined
 and summarized into a phase diagram of the symmetry.
It is revealed
 that there is an effective threshold in $\mu$
 for the appearance of
 the $d$-wave superconductivity
 regardless of the strength of the $d$-channel interaction.
It is also shown
 that,
 although the $s$-wave superconductivity
 can occur for any value of $\mu$
 if the $s$-channel interaction is mixed sufficiently,
 the $d$-wave superconductivity
 has the advantage of achieving a high $T_{c}$
 over the $s$-wave one,
 once it occurs.
\end{abstract}

\noindent

\begin{center}
  mixed interaction of $s$- and $d$-wave, band-filling, high-$T_{c}$
\end{center}

\vspace{2in}

The corresponding author: Katsuji Sakai\\
\indent
TEL: +81-3-3260-4271 ext.2425,
 FAX: +81-3-3260-4772,\\
\indent
e-mail: sakai@grad.ap.kagu.sut.ac.jp\\

\newpage

\setcounter{page}{1}
\pagestyle{plain}


\section{Introduction}

Since the discovery
 of the high $T_{c}$ copper oxide superconductors (the cuprates),
 much attention has been directed
 to the symmetry of the Cooper-pair
 in the cuprates.
This is
 partly because we have not yet found a decisive origin
 of the superconductivity in cuprates,
 and it is expected
 that the decision of the symmetry of the superconducting order parameter
 would illuminate this issue.
Among candidates of the pairing symmetry,
 much of discussions have been focused
 on the $s$-wave and the $d$-wave symmetry,
 and it has been a central interest
 which of the two occurs really in cuprates.
A lot of experimental investigations
 have been done
 with regard to this subject
\cite{experiments}.
Their results, however, have not focused
 decisively on one side of the two.
Although recent experiments seem to suggest
 dominantly the $d_{x^{2}-y^{2}}$ symmetry,
 the results are still controversial
 even in the new experiments
 using the Josephson-junction
 which intend to measure
 the phase of the order parameter;
 some
 led to results
 taking the side of the $d_{x^{2}-y^{2}}$ symmetry
\cite{d-wave}
 while others showed results
 in favor of the $s$-wave symmetry
\cite{s-wave}.

On the theoretical side,
 there seems to be a general feeling
 that the realization
 of the $d_{x^{2}-y^{2}}$ pairing state in the cuprates
 rules out
 the electron-phonon interaction
 as the origin of the superconductivity
 while that of the $s$-wave pairing state
 rules out
 the antiferromagnetic(AFM) spin-fluctuation exchange mechanism
 and supports the electron-phonon interaction
 as a main origin.
Of course,
 there have been some works
 against this common feeling.
A recent work
\cite{PRL74-2303}, for example,
 has shown
 that, even if the AFM spin-fluctuation
 in a CuO$_{2}$ layer
 is responsible for the superconductivity in the cuprates,
 the $s$-wave pairing state
 can be stabilized
 for a bilayer system of CuO$_{2}$ planes
 by virtue of the existence of strong AFM spin correlations
 between the constituent CuO$_{2}$ layers.
On the other hand,
 some work has predicted
 the $d$-wave superconductivity to occur
 even due to the electron-phonon interaction
 by starting from an effective single-band Hubbard-type Hamiltonian
 which includes
 both the strong electron-electron repulsion
 and the electron-phonon coupling due to oxygen vibrational modes
 in CuO$_{2}$ planes
\cite{PRB51-3840}.
An other work showed also
 that the electron-phonon interaction
 leads to the $d$-wave superconductivity
 based on the ``two-story-house-model"
 which assumes
 the antiferromagnetically correlated Cu spins
 as a background of itinerant carriers
\cite{PRL77-723}.
These few examples
 tell us
 that the pairing symmetry
 does not correspond uniquely to the pairing mechanism
 and their realization depends also on the model Hamiltonian
 which is assumed to describe the electronic state;
 one-band or multi-band,
 one-layer or coupled-layers,
 the strong electronic correlation
 being taken into account or not,
 background AFM spin correlation
 being considered or not
 and so on.
The band-filling
 as well as the type of the interaction
 responsible for the superconductivity
 clearly affects
 what kind of the pairing symmetry is realized.

In such an entangled situation,
 it would be necessary for us
 to simplify the problem and to start with minimal ingredients.
As a first step,
 we restrict ourselves to a two-dimensional CuO$_{2}$ layer
 but do
 not restrict ourselves
 to any particular mechanism of the superconductivity.
To start with,
 we address the following
 simple questions:
When a pairing interaction composed of the $s$-wave and the $d$-wave channel
 is introduced into the carriers in a CuO$_{2}$ plane,
 what kind of pairing symmetry of the superconducting state
 is realized?
How does this result
 depend on the band-filling
 and the mixing ratio between different channels in the interaction?

This
 question, however, is not quite new.
Lenck et al.,
 on the basis of almost the same idea
 but specifying the origins of the interactions
 to the exchanges of phonons and AFM spin-fluctuations,
 have examined this problem
 by solving the Eliashberg equations
 for carriers described by a two-dimensional tight-binding band
\cite{JLTP80-269}.
They gave
 results, however,
 only for several values of the mixing ratio and the band-filling.
The main purpose of this paper
 is to make a $\mu-s$ phase diagram
 for the pairing symmetry surviving in a two-dimensional superconductor,
 where $s$ denotes the mixing parameter of the interaction of $s$-wave symmetry
 and $\mu$ is the chemical potential
 which is related to the band-filling of the carriers.
Assuming
 a simple two-dimensional nearest-neighbor tight-binding band,
 we solve two-dimensional Eliashberg equations
 by moving thoroughly the mixing parameter $s$ and the chemical potential $\mu$.
Such a detailed ${\mu}-s$ phase diagram for the surviving pairing symmetry
 has not yet been given by others.
The overall tendency of our results
 coincides with the previous one\cite{JLTP80-269}
 given by others in a specified region of $\mu$.
We calculate, further,
 the superconducting critical temperature $T_{c}$
 in the realized symmetry state
 as a function of the mixing parameter $s$ and the chemical potential $\mu$.

The plan of this paper is as follows.
In Section 2,
 our model for the interaction is introduced
 and a brief survey of our framework of the calculation is given.
The 
 essential ingredients for the Eliashberg equations
 are also presented.
In Section 3,
 the resulting phase-diagrams for the pairing symmetry
 are discussed.
Characteristic difference
 between the behaviors of $T_{c}$ in different symmetry states
 (the $s$-wave and the $d$-wave superconducting state)
 is also pointed out.
In the last section,
 the results of this paper are summarized.


\section{Formalism}

We start
 from writing down the two-dimensional Eliashberg equations
 on the imaginary-axis
 for carriers with a band energy $\epsilon_{\mbox{\boldmath$k$}}$
 in the form of coupled equations
 for the renormalization function $Z(\mbox{\boldmath$k$}, i\omega_{n})$,
 the normal self-energy $\chi(\mbox{\boldmath$k$}, i\omega_{n})$
 and the anomalous self-energy $\phi(\mbox{\boldmath$k$}, i\omega_{n})$
:
\begin{equation}
[1-Z(\mbox{\boldmath$k$},i\omega_{n})] \omega_{n}  =  -\frac{1}{\beta} \sum_{\mbox{\boldmath$k'$},n'}
\frac{\omega_{n'} Z(\mbox{\boldmath$k'$},i\omega_{n'} )\lambda(\mbox{\boldmath$k$},\mbox{\boldmath$k'$},n-n')}
     {{\omega_{n'} }^{2}Z^{2}( \mbox{\boldmath$k'$},i\omega_{n'}) +
          {\phi}^{2}( \mbox{\boldmath$k'$},i\omega_{n'}) +
[\epsilon_{\mbox{\boldmath$k'$}} + \chi(\mbox{\boldmath$k'$},i\omega_{n'})]^2}
\: ,
\end{equation}
\begin{equation}
 \phi( \mbox{\boldmath$k$},i\omega_{n} )  =  \frac{1}{\beta} \sum_{ \mbox{\boldmath$k'$},n'}
 \frac{\phi(\mbox{\boldmath$k'$},i\omega_{n'} )
[\lambda(\mbox{\boldmath$k$},\mbox{\boldmath$k'$},n-n') - V(\mbox{\boldmath$k$}-\mbox{\boldmath$k'$})]}
     { {\omega_{n'} }^{2}Z^{2}( \mbox{\boldmath$k'$},i\omega_{n'} ) +
          {\phi}^{2}( \mbox{\boldmath$k'$},i\omega_{n'} ) +
[\epsilon_{\mbox{\boldmath$k'$}} + \chi(\mbox{\boldmath$k'$},i\omega_{n'} )]^2}
\: ,
\end{equation}
\begin{equation}
\chi( \mbox{\boldmath$k$},i\omega_{n} )  =  -\frac{1}{\beta} \sum_{ \mbox{\boldmath$k'$},n'}
\frac{[\epsilon_{\mbox{\boldmath$k'$}} + \chi(\mbox{\boldmath$k'$},i\omega_{n'})]
\lambda(\mbox{\boldmath$k$},\mbox{\boldmath$k'$},n-n')}
{ {\omega_{n'} }^{2}Z^{2}( \mbox{\boldmath$k'$},i\omega_{n'} ) +
     {\phi}^{2}( \mbox{\boldmath$k'$},i\omega_{n'} ) +
[\epsilon_{\mbox{\boldmath$k'$}} + \chi(\mbox{\boldmath$k'$},i\omega_{n'})]^2}
\: ,
\end{equation}
where wave vectors $\mbox{\boldmath$k$}$ and $\mbox{\boldmath$k'$}$
 lie in the a-b plane(CuO$_{2}$ plane)
 and $\omega_{n}=(2n+1){\pi}/\beta$
 with $\beta=1/k_{B}T$.
Here the band energy $\epsilon_{\mbox{\boldmath$k$}}$
 is measured
 with respect to the bare band structure chemical potential $\mu$.
The Coulomb interaction $V(\mbox{\boldmath$k$}-\mbox{\boldmath$k'$})$
 will be neglected,
 for simplicity, in the following calculation.
In the usual phonon-mediated theory,
 the interaction kernel
 $\lambda(\mbox{\boldmath$k$}, \mbox{\boldmath$k'$}, n-n')$
 is written
 by means of the electron-phonon coupling constant
 $\bar{g}_{\mbox{\boldmath$k$},\mbox{\boldmath$k'$},\lambda}$
 and the propagator
 $D_{\lambda}(\mbox{\boldmath$k$}-\mbox{\boldmath$k'$},i\omega_{n}-i\omega_{n'})$
 of phonons with polarization $\lambda$
 as the following expression,
\begin{equation}
\lambda(\mbox{\boldmath$k$},\mbox{\boldmath$k'$},n-n')
\equiv
-\sum_{\lambda}|\overline{g}_{kk'\lambda}|^2
D_\lambda(\mbox{\boldmath$k$}-\mbox{\boldmath$k$}',i\omega_{n} - i\omega_{n'})
\: ,
\end{equation}
and this is expressed
 further in terms of the electron-phonon spectral function
 $\alpha^{2}F(\mbox{\boldmath$k$}, \mbox{\boldmath$k'$},\Omega)$ as
\begin{equation}
   \lambda(\mbox{\boldmath$k$},\mbox{\boldmath$k'$},n) = \int_{0}^{\infty} d\Omega\
   \frac{2\Omega \alpha^2 F(\mbox{\boldmath$k$},\mbox{\boldmath$k'$},\Omega)}
   {(2n\pi k_B T)^2 + \Omega^2}
                \: .
\end{equation}
In this case,
 $\alpha^{2}F(\mbox{\boldmath$k$}, \mbox{\boldmath$k'$},\Omega)$
 is positive definite
 for any value of wave vectors
 $\mbox{\boldmath$k$}$,$\mbox{\boldmath$k'$}$ and the phonon energy $\Omega$.
But we will consider, from now on,
 a general electron-boson interaction, therefore,
 this positive definiteness
 of $\alpha^{2}F(\mbox{\boldmath$k$},\mbox{\boldmath$k'$},\Omega)$
 is not necessarily maintained.
Since we assume
 the square symmetry of the CuO$_{2}$ plane
 by neglecting the anisotropy in the a-b plane,
 the general interaction spectral function
 $\alpha^{2}F(\mbox{\boldmath$k$},\mbox{\boldmath$k'$},\Omega)$
 is expanded
 by means of the basis functions $\Psi_{i}(\mbox{\boldmath$k$})$
 of the point group $D_{4h}$.
Here we assume, for simplicity,
 that the mixing interaction
 consists of the $s$-wave channel and the $d$-wave channel.
Then, in our model,
 the interaction spectral function is expanded
 in terms of basis functions
 $\Psi_{s}(\mbox{\boldmath$k$})=1$
 and
 $\Psi_{d}(\mbox{\boldmath$k$})
=\cos(\mbox{\boldmath$k_{x}$})-\cos(\mbox{\boldmath$k_{y}$})$
 as follows;
\begin{equation}
{\alpha}^{2}F(\mbox{\boldmath$k$},\mbox{\boldmath$k'$},\Omega)
=\overline{{\alpha}^{2}F_{s}(\Omega)}\Psi_{s}(\mbox{\boldmath$k$})\Psi_{s}(\mbox{\boldmath$k'$})
+\overline{{\alpha}^{2}F_{d}(\Omega)}\Psi_{d}(\mbox{\boldmath$k$})\Psi_{d}(\mbox{\boldmath$k'$})
\: .
\end{equation}
Here it should be noted
 that $\overline{{\alpha}^{2}F_{i}(\Omega)}$ (i = s, d)
 are projections of the interaction spectral function
 into the $s$-channel and the $d$-channel.
Therefore, $\lambda(\mbox{\boldmath$k$}, \mbox{\boldmath$k'$}, n-n')$
 can be also written as
\begin{equation}
\lambda(\mbox{\boldmath$k$},\mbox{\boldmath$k'$},n)
=\lambda_{s}(n)\Psi_{s}(\mbox{\boldmath$k$})\Psi_{s}(\mbox{\boldmath$k'$})
+\lambda_{d}(n)\Psi_{d}(\mbox{\boldmath$k$})\Psi_{d}(\mbox{\boldmath$k'$})
\: ,
\end{equation}
where
\begin{equation}
   \lambda_{i}(n) = \int_{0}^{\infty} d\Omega\
   \frac{2\Omega \overline{{\alpha}^2 F_{i}(\Omega)}}
   {(2n\pi k_B T)^2 + \Omega^2}
 \qquad (i=s,d)\: .
\end{equation}
%
Consequently, 
 from the structure of the Eliashberg equations
 (1)$\sim$(3),
 we can easily see
 that the functions
 $Z(\mbox{\boldmath$k$}, i\omega_{n})$,
 $\chi(\mbox{\boldmath$k$}, i\omega_{n})$
 and $\phi(\mbox{\boldmath$k$}, i\omega_{n})$
 are also expanded in a following way;

\begin{equation}
Z(\mbox{\boldmath$k$},i{\omega}_{n})
=Z_{s}(i{\omega}_{n})\Psi_{s}(\mbox{\boldmath$k$})
+Z_{d}(i{\omega}_{n})\Psi_{d}(\mbox{\boldmath$k$})
\: ,
\end{equation}
\begin{equation}
\phi(\mbox{\boldmath$k$},i{\omega}_{n})
={\phi}_{s}(i{\omega}_{n})\Psi_{s}(\mbox{\boldmath$k$})
+{\phi}_{d}(i{\omega}_{n})\Psi_{d}(\mbox{\boldmath$k$})
\: ,
\end{equation}
\begin{equation}
\chi(\mbox{\boldmath$k$},i{\omega}_{n})
={\chi}_{s}(i{\omega}_{n})\Psi_{s}(\mbox{\boldmath$k$})
+{\chi}_{d}(i{\omega}_{n})\Psi_{d}(\mbox{\boldmath$k$})
\: .
\end{equation}
These expansions reduce the Eliashberg equations
 to the coupled equations
 for the functions
 $Z_{i}(i\omega_{n})$,
 $\chi_{i}(i\omega_{n})$
 and $\phi_{i}(i\omega_{n})$
 with $i = s, d$;

\begin{equation}
Z_{s}(i\omega_{n}) = 1+\frac{1}{\omega_{n}\beta}\sum_{k', n'}
\frac{\omega_{n'}\lambda_{s}(n-n')
[Z_{s}(i\omega_{n'})\Psi_{s}(\mbox{\boldmath$k'$})
+Z_{d}(i\omega_{n'})\Psi_{d}(\mbox{\boldmath$k'$})]
\Psi_{s}(\mbox{\boldmath$k'$})}
{A(\mbox{\boldmath$k'$}, i\omega_{n'})}
\: ,
\end{equation}
\begin{equation}
Z_{d}(i\omega_{n}) = \frac{1}{\omega_{n}\beta}\sum_{k', n'}
\frac{{\omega_{n'}}{\lambda_{d}(n-n')}
[Z_{s}(i\omega_{n'})\Psi_{s}(\mbox{\boldmath$k'$})
+Z_{d}(i\omega_{n'})\Psi_{d}(\mbox{\boldmath$k'$})]
\Psi_{d}(\mbox{\boldmath$k'$})}
{A(\mbox{\boldmath$k'$}, i\omega_{n'})}
\: ,
\end{equation}
\begin{equation}
\phi_{s}(i\omega_{n}) = \frac{1}{\beta}\sum_{k', n'}
\frac{\lambda_{s}(n-n')
[\phi_{s}(i\omega_{n'})\Psi_{s}(\mbox{\boldmath$k'$})
+\phi_{d}(i\omega_{n'})\Psi_{d}(\mbox{\boldmath$k'$})]
\Psi_{s}(\mbox{\boldmath$k'$})}
{A(\mbox{\boldmath$k'$}, i\omega_{n'})}
\: ,
\end{equation}
\begin{equation}
\phi_{d}(i\omega_{n}) = \frac{1}{\beta}\sum_{k', n'}
\frac{\lambda_{d}(n-n')
[\phi_{s}(i\omega_{n'})\Psi_{s}(\mbox{\boldmath$k'$})
+\phi_{d}(i\omega_{n'})\Psi_{d}(\mbox{\boldmath$k'$})]
\Psi_{d}(\mbox{\boldmath$k'$})}
{A(\mbox{\boldmath$k'$}, i\omega_{n'})}
\: ,
\end{equation}
\begin{equation}
\chi_{s}(i\omega_{n}) = - \frac{1}{\beta}\sum_{k', n'}
\frac{\lambda_{s}(n-n')
[\chi_{s}(i\omega_{n'})\Psi_{s}(\mbox{\boldmath$k'$})
+\chi_{d}(i\omega_{n'})\Psi_{d}(\mbox{\boldmath$k'$})
+\epsilon_{\mbox{\boldmath$k'$}}]
\Psi_{s}(\mbox{\boldmath$k'$})}
{A(\mbox{\boldmath$k'$}, i\omega_{n'})}
\: ,
\end{equation}
\begin{equation}
\chi_{d}(i\omega_{n}) = - \frac{1}{\beta}\sum_{k', n'}
\frac{\lambda_{d}(n-n')
[\chi_{s}(i\omega_{n'})\Psi_{s}(\mbox{\boldmath$k'$})
+\chi_{d}(i\omega_{n'})\Psi_{d}(\mbox{\boldmath$k'$})
+\epsilon_{\mbox{\boldmath$k'$}}]
\Psi_{d}(\mbox{\boldmath$k'$})}
{A(\mbox{\boldmath$k'$}, i\omega_{n'})}
\: ,
\end{equation}
where
\begin{eqnarray}
A(\mbox{\boldmath$k'$}, i\omega_{n'}) & \equiv &
{\omega_{n'} }^2
[Z_{s}(i\omega_{n'})\Psi_{s}(\mbox{\boldmath$k'$})
+Z_{d}(i\omega_{n'})\Psi_{d}(\mbox{\boldmath$k'$})]^2
+[\phi_{s}(i\omega_{n'})\Psi_{s}(\mbox{\boldmath$k'$})
+\phi_{d}(i\omega_{n'})\Psi_{d}(\mbox{\boldmath$k'$})]^2
\nonumber \\
  &   &
+[\epsilon_{\mbox{\boldmath$k'$}}
+\chi_{s}(i\omega_{n'})\Psi_{s}(\mbox{\boldmath$k'$})
+\chi_{d}(i\omega_{n'})\Psi_{d}(\mbox{\boldmath$k'$})]^2
\: .
\end{eqnarray}
In our model,
 the itinerant electrons
 are assumed
 to be described
 by the nearest-neighbor tight-binding band
\begin{equation}
\epsilon_{\mbox{\boldmath$k$}}=2\bar{t}\ [2-\cos(k_x)-\cos(k_y)-\bar{\mu}]
\: ,
\end{equation}
where $\bar{t}$ is the nearest-neighbor hopping energy
 (taken as 170[meV]
 in later calculations)
 and $\bar{\mu}$ is the chemical potential
 normalized with $2\bar{t}$.
 These self-consistent coupled equations
 (12)$\sim$(17)
 are solved numerically.
There,
 the wave number summations
 over the square Brillouin zone $(-\pi \leq k_{i} < \pi ;i=x,y)$
 are carried out numerically.
The expansion of $\alpha^{2}F(\mbox{\boldmath$k$},\mbox{\boldmath$k'$},\Omega)$
 in such a way
 as given in eq.(6)
 has already been used
 in several works
\cite{JLTP80-269,PRB41-7289}.
In some work,
 however, the wave number summation
 was inverted into an angular integration in k-space
 assuming the isotropic Fermi surface,
 and therefore a full summation over the first Brillouin zone
 was avoided
\cite{PRB41-7289}.

A final important ingredient
 to be provided in the calculation
 is the form of $\overline{\alpha^{2}F_{i}(\Omega)}$
 appearing in eq.(6).
Even after $\alpha^{2}F(\mbox{\boldmath$k$},\mbox{\boldmath$k'$},\Omega)$
 is expanded
 as in eq.(6),
 a selective oppotunity
 is still left us
 in the choice
 of the functional form
 for $\overline{\alpha^{2}F_{s}(\Omega)}$
 and $\overline{\alpha^{2}F_{d}(\Omega)}$ .
We may, of course,
 employ properly
 two model functions for them
 different from each other.
The numerical results
 will suffer from a slight modification
 dependently on the choice
 of the functional form for them.
To investigate closely
 in what manner
 they are affected
 by the detailed form of $\overline{\alpha^{2}F_{i}(\Omega)}$
 is a problem in itself
 worth being studied,
 however,
 we postpone it
 on the next occasion.
For our own purpose of this paper,
 we here make a simple choice for them
 such that both of $\overline{\alpha^{2}F_{s}(\Omega)}$
 and $\overline{\alpha^{2}F_{d}(\Omega)}$
 have the same functional form
 except for multiplicative constants;
 that is,
 $\overline{\alpha^{2}F_{s}(\Omega)}=
s\hspace{1mm}\overline{\alpha^{2}F(\Omega)}$
 and
 $\overline{\alpha^{2}F_{d}(\Omega)}=
d\hspace{1mm}\overline{\alpha^{2}F(\Omega)}$
 under the condition $s+d=1$.
%
Then we can concentrate on a problem
 what kind of the symmetry of Cooper-pair
 can appear
 when the ratio of the intensity
 between the $s$-wave and the $d$-wave component
 in the electron-boson spectral function
 is varied continuously
 and the band-filling (equivalently, the chemical potential)
 is done samely.
The function $\overline{\alpha^{2}F(\Omega)}$,
 in this case,
 is quite different
 in its meaning
 from the electron-phonon spectral function $\alpha^{2}F(\Omega)$
 in the conventional Eliashberg equations.
As for the functional form of $\overline{\alpha^{2}F(\Omega)}$,
 however,
 we employ a function
 which was used in the previous work
 being expected
 to model the spectral function in Bismuth-based cuprates
\cite{SSC76-1189}.
The functional form
 of this $\overline{\alpha^{2}F(\Omega)}$
 is shown in Fig.1.
It should be mentioned, however,
 that we tentatively employ
 this model function
 merely from a reason
 that it has comparatively wide spectral distribution
 as the electronic band has.
One may suspect
 that such a choice of $\overline{\alpha^{2}F_{i}(\Omega)}$
 makes a considerable influence
 on the numerical results
 and the complex form of $\overline{\alpha^{2}F(\Omega)}$
 as well as the minute phonon structures of it
 will affect the conclusions.
We will make comments briefly
 on this point in Section 3.
%
%
%
%


\section{Phase diagrams for the pairing symmetry}

For many sets of parameters $(\bar{\mu},s)$,
 the reduced chemical potential $\bar{\mu}$
 and the mixing parameter $s$ of the $s$-channel interaction,
 the Eliashberg equations (12)$\sim$(17) are solved
 at various temperatures.
In the calculations,
 a model function $\overline{\alpha^{2}F(\Omega)}$
 is used
 which is decided
 to give $T_{c}=80[K]$
 for values of parameters $(\bar{\mu},s)=(1.5,1.0)$.

At each fixed temperature $T$,
 for a given set of values $(\bar{\mu},s)$
 we examine
 whether the coupled equations (12)$\sim$(17)
 have non-vanishing solutions
 of the anomalous self-energy $\phi_{j}(i\omega_{n})$ or not.
If we find out the solutions $\phi_{s}(i\omega_{n})$
 to be non-vanishing for all $n$,
 then we recognize
 that for a given set of $(\bar{\mu},s)$
 the state is superconducting
 with the $s$-wave pairing symmetry at this temperature $T$.
If we find out, on the other hand,
 the solutions $\phi_{d}(i\omega_{n})$
 to be non-vanishing,
 we recognize
 that the state is superconducting
 with the $d$-wave pairing symmetry at the same temperature $T$.
From the symmetry consideration
 of the form of $\alpha^{2}F(\mbox{\boldmath$k$},\mbox{\boldmath$k'$},\Omega)$
 in eq.(6),
 it is understood
 that the occurrence of the pairing state with a mixed symmetry
 is not possible in this case.
If we find
 that both of the solutions
 $\phi_{s}(i\omega_{n})$ and $\phi_{d}(i\omega_{n})$
 vanish for all $n$,
 we understand
 the state to be normal
 for such values of $(\bar{\mu},s)$
 at the temperature $T$.
At various temperatures,
 in this way,
 by changing the values of the parameters $(\bar{\mu},s)$
 we specify the pairing symmetry of the solutions
 one after the other.
We exhibit the resulting diagram
 in the $\bar{\mu}-s$ plane
 for the specification of states
 realized at $T$=20[K] and 40[K]
 in Fig.2(a) and (b), respectively.
There,
 circles and triangles
 indicate points $(\bar{\mu},s)$
 for which the states are superconducting
 with the $s$-wave and the $d$-wave pairing symmetry, respectively,
 and squares denote points
 for which the states are normal
 at the temperatures.
For clarity,
 the points
 only in the close proximity
 of the boundary lines
 between regions
 specified by different symmetries
 are plotted
 by picking up
 from a larger number of points
 examined.

We can easily understand
 that the boundary line
 between the normal region
 and any of the superconducting regions
 (of $s$-wave or $d$-wave)
 represents just a ``equi$-T_{c}$" line,
 for the points $(\bar{\mu},s)$
 on which the superconductivity
 (of the $s$-wave or the $d$-wave pairing)
 can be produced
 with the critical temperature
 $T_{c}$=20[K] in Fig.2(a) and $T_{c}$=40[K] in Fig.2(b), respectively.
In these diagrams,
 actual calculations are made
 for $\bar{\mu}$
 in a region 0.25$\leq\bar{\mu}$,
 because the reliability
 of the calculations
 becomes worse
 in the close proximity
 of $\bar{\mu}$=0:
When $\bar{\mu}$ is close to zero,
 the Fermi energy
 becomes much less
 than the spectral width of $\overline{\alpha^{2}F(\Omega)}$,
 so that the Eliashberg calculation
 loses its meaning.

Combining the results at $T$=20[K] and 40[K]
 with additional ones at $T$=5[K] and 60[K],
 we make up
 a phase diagram of the pairing symmetry
 in the $\bar{\mu}-s$ plane,
 which is depicted in Fig.3.
The region hatched by vertical dotted lines
 is here after called
 that of the ``$S$-phase"
 where sets of the coordinates $(\bar{\mu},s)$
 can produce $s$-wave superconductivities.
The region hatched by horizontal dotted lines
 is of the ``$D$-phase",
 which has a corresponding meaning
 same as in the case of the term ``$S$-phase".
A thick solid line
 represents
 the boundary between regions of different phases.
Several thin solid lines drawn together
 are ``equi-$T_{c}$" lines;
 the value of $T_{c}$
 is indicated
 on each side of them.
The boundary line
 has not been determined eventually
 in a region
 of the phase diagram
 where $\bar{\mu}<$0.5 and $s\simeq$0,
 because we could not solve
 the Eliashberg equations
 at very low temperatures
 below 5[K]
 due to the limitations
 of the computational time.
The boundary
 drawn in a thick dotted line
 in this region
 is, therefore,
 only a guess
 from the behaviors
 of the solutions
 of the Eliashberg equations
 at 5[k]
 for some points of $(\bar{\mu},s)$
 in this region.

Looking in this phase diagram in Fig.3,
 we first notice three points:
 (i)The $d$-wave superconductivity
 can hardly occur for small $\bar{\mu}$,
 even if the $d$-channel interaction
 which causes the $d$-wave pairing
 is mixed strongly.
It seems
 that there is a lower bound for $\bar{\mu}$
 in order for the $d$-wave superconductivity
 to be able to appear.
(ii)To the contrary,
 the $s$-wave superconductivity can occur
 regardless of the value of $\bar{\mu}$.
Even for large $\bar{\mu}$
 near $\bar{\mu}$=2 (half-filling case),
 it can occur
 if the $s$-channel interaction
 is sufficiently strong
 compared with the $d$-channel one,
 i.e. $s/d>$4.
(iii)The boundary curve
 between the $S$-phase and the $D$-phase region
 has a positive gradient
 and goes into the upper-half
 of the phase diagram ($s>$0.5)
 when $\bar{\mu}\geq$1.25.
This means
 that even if the $s$-channel interaction
 is mixed largely,
 the $d$-wave superconductivity
 can preferably occur
 when $\bar{\mu}$
 is
 sufficiently
 close
 to the half-filling value 2.
And this preference progresses
 increasingly as $\bar{\mu}$ increases.
Reversely,
 the $s$-wave superconductivity
 occurs preferably
 as $\bar{\mu}$ becomes
 smaller than 1.25.
This preference progresses
 much more rapidly
 as $\bar{\mu}$ decreases
 than the preference of the $d$-wave superconductivity
 does as $\bar{\mu}$ increases
 in the case of the $d$-wave,
 since the gradient of the boundary curve
 is steeper
 for smaller $\bar{\mu}$.

These characteristic features
 of the phase diagram
 are well expected
 and agree with common feelings.
From the calculational point of view,
 this can be understood as follows.
By looking in Eq.(15),
 for example,
 of coupled equations (12)$\sim$(17),
 we can see
 that, in order for $\phi_{d}(i\omega_{n})$
 to survive
 rather than $\phi_{s}(i\omega_{n})$ to do,
 it is necessary
 for the value of $\Psi_{d}(\mbox{\boldmath$k'$})^{2}$
 in the numerators of the summand
 on the right hand side of the equation
 to be large compared with the value of $\Psi_{s}(\mbox{\boldmath$k'$})^{2}$
 around the Fermi surface(line).
There, for small $\bar{\mu}$,
 $\Psi_{d}(\mbox{\boldmath$k'$})^{2}$
 is much smaller than $\Psi_{s}(\mbox{\boldmath$k'$})^{2}$(=1),
 and the effective coupling strength
 of the $d$-channel interaction
 proportional to $\Psi_{d}(\mbox{\boldmath$k'$})^{2}$
 is entirely overwhelmed
 by that of the $s$-channel one.
Therefore,
 the $d$-wave superconductivity
 hardly appears
 in this case.
As the value of $\bar{\mu}$ increases
 and approaches to the half-filling value 2,
 the maximum value of $\Psi_{d}(\mbox{\boldmath$k'$})^{2}$
 around the Fermi surface approaches to 4,
 therefore the average value of $\Psi_{d}(\mbox{\boldmath$k'$})^{2}$
 around the Fermi surface
 becomes much higher
 than the value of $\Psi_{s}(\mbox{\boldmath$k'$})^{2}$(=1).
This produces
 effectively large coupling strength
 of the $d$-channel interaction,
 which causes
 the preference of the $d$-wave superconductivity
 in the proximity of $\bar{\mu}$=2.

From Fig.3,
 we can deduce also the behavior of $T_{c}$
 as a function of $\bar{\mu}$ and $s$.
We first notice
 that ``equi-$T_{c}$" curves
 are
 distributed densely
 in the $D$-phase region
 while those in the $S$-phase region
 are sparse.
This means
 that if the point $(\bar{\mu},s)$,
 starting from a some point
 on the boundary line
 between the different phase regions,
 moves away from it
 along the normal directions
 of respective ``equi-$T_{c}$" lines
 on both sides of the boundary,
 the value of $T_{c}(\bar{\mu},s)$
 increases much more rapidly on the $D$-phase side
 than on the $S$-phase side.
In the $D$-phase region
 $T_{c}(\bar{\mu},s)$
 would have the highest value at $(\bar{\mu},s)$=(2,0),
 which would be much larger
 than the highest value at $(\bar{\mu},s)$=(2,1)
 in the $S$-phase region.
Secondly,
 from an inspection of the phase diagram,
 we can find
 the dependence of $T_{c}$
 on the mixing parameter $s$
 for the fixed value of $\bar{\mu}$.
In Fig.4,
 the curves of $T_{c}(\bar{\mu},s)$ versus $s$
 are shown
 for several fixed values of $\bar{\mu}$.
There,
 each $T_{c}-s$ curve
 has a kink-like minimum point at $s=s_{0}$
 which is the ordinate
 of the crossing point $(\bar{\mu},s_{0})$
 of the line of constant $\bar{\mu}$
 and the boundary line
 between the $S$-phase and the $D$-phase region
 in the phase diagram shown in Fig.3.
In
 the
 curves for $\bar{\mu}$=1.5 and 0.8,
 we notice
 that $T_{c}$ increases much more rapidly
 as the $d$-channel mixing parameter $d=1-s$
 increases from $d_{0}=1-s_{0}$
 than it does
 as the $s$-channel mixing parameter $s$
 increases from $s_{0}$.
The magnitude of the gradient of the $T_{c}-s$ curve
 in the $D$-phase region$(s<s_{0})$
 depends on the value of fixed $\bar{\mu}$;
 the gradient becomes steeper
 for larger values of $\bar{\mu}$.
On the other hand,
 in the $S$-phase region$(s>s_{0})$
 the gradient is almost independent of the value of $\bar{\mu}$.
This larger gain of $T_{c}$
 in the $D$-phase
 than in the $S$-phase
 for a same amount of increase
 of respective parameters $d$ and $s$
 can be understood
 as follows.
In the $D$-phase region,
 in addition to the predominance
 of the $d$-channel effective coupling
 compared to the $s$-channel one,
 the reduction of $T_{c}$
 by the damping effect
 owing to the renormalization function $Z(\mbox{\boldmath$k$},i\omega_{n})$
 is expected
 to be small.
Because $Z_{d}(i\omega_{n})$ vanishes
 due to its $d$-wave symmetry
 even for large value of $d$
 and $Z_{s}(i\omega_{n})$
 is small
 due to smallness of $s$
 in the $D$-phase region,
 the renormalization function $Z(\mbox{\boldmath$k$},i\omega_{n})$
 is small
 as a whole
 in the $D$-phase region.
Therefore, $T_{c}$ is much more sensitive
 to the change of electron-boson coupling
 in $D$-phase
 than in $S$-phase.
This characteristic behavior of $T_{c}(\bar{\mu},s)$
 for fixed $\bar{\mu}$ values
 is also guessed
 from the estimation by eye
 the directional derivative of $T_{c}(\bar{\mu},s)$
 along the line of constant $\bar{\mu}$
 in the phase diagram given in Fig.3:
The directional derivatives
 along each line of constant $\bar{\mu}$
 changes its sign
 and is discontinuous
 at the point $(\bar{\mu},s_{0})$
 on the boundary line.
The density of the ``equi-$T_{c}$" lines
 along the line of constant $\bar{\mu}$
 is much higher in the $D$-phase region$(s<s_{0})$
 than in the $S$-phase region$(s>s_{0})$.
And this high density of the ``equi-$T_{c}$" line
 along the line of constant $\bar{\mu}$
 in the $D$-phase region
 progresses increasingly
 as the constant $\bar{\mu}$ value
 becomes larger
 while it scarcely changes
 in the $S$-phase region.
Thirdly,
 we can see
 that, for a fixed mixing parameter $s$,
 $T_{c}(\bar{\mu},s)$ increases
 monotonously as $\bar{\mu}$ does,
 but the gradient of the $T_{c}-\bar{\mu}$ curve
 in the $D$-phase region
 is much larger
 than that in the $S$-phase region.
This larger gradient
 in the $D$-phase 
 is also explained
 from the strong enhancement
 of the effective electron-boson coupling strength
 by the increase of $\bar{\mu}$
 in the $D$-phase.

Finally we should mention
 that the absolute value of $T_{c}$ itself
 does not have important meaning
 in
 this explanation of the phase diagram.
Since in our calculation
 we employ the model function $\overline{\alpha^{2}F(\Omega)}$
 having a certain multiplicative constant
 tentatively chosen,
 the $T_{c}$ value can be modified
 in any way
 by this multiplicative constant.
However,
 the characteristic feature
 of the distribution
 of the ``equi-$T_{c}$" lines
 is expected
 to remain almost unaltered
 unless we use a meaninglessly large multiplicative constant.
Although the value of $T_{c}$
 beside each ``equi-$T_{c}$" line
 would be scaled
 by this multiplicative constant,
 the boundary line
 which divides
 the different phase regions
 would remain almost the same.

Here we would like to make a comment
 on the effect
 of the minute structures
 in the model functions $\overline{\alpha^{2}F_{i}(\Omega)}$
 on the numerical results:
To put it concisely,
 the minute structures
 in $\overline{\alpha^{2}F_{i}(\Omega)}$
 do not give
 any significant and qualitative change
 to the feature
 of the results stated above.
Since, in the Eliashberg equations (12)$\sim$(17),
 $\overline{\alpha^{2}F_{i}(\Omega)}$
 plays its role
 through the $n$-dependence of $\lambda_{i}(n)$
 given in eq.(8),
 the minute structures of $\overline{\alpha^{2}F_{i}(\Omega)}$
 are smeared out
 because of the Lorenzian
 in the integrand
 of the right-hand side in eq.(8)
 so that any serious effect
 is not produced
 by their presence.
%
If we employ another forms for $\overline{\alpha^{2}F_{i}(\Omega)}$,
 of course,
 results stated above
 may be modified
 more or less quantitatively:
At finite temperatures,
 the boundary curve
 between the normal region and the superconducting region
 will be modified slightly.
The boundary curve
 between the $S$-phase and the $D$-phase region
 wil be  also modified in the same way;
 we have checked this
 by making additional calculations with a cutoff Lorenzian-type
 of the model function for $\overline{\alpha^{2}F(\Omega)}$.
But, in this paper,
 we do not discuss
 details of the effect
 due to the change
 of the functional form for $\overline{\alpha^{2}F(\Omega)}$,
 because it gives only minor effects
 to the results
 in the context
 of our primary concern
 as well as it is beyond the purpose of this paper.


\section{Conclusions}

We have solved the two-dimensional Eliashberg equations
 employing an electron-boson interaction
 which has
 the $s$-wave channel and the $d$-wave channel.
The electronic state
 is assumed to be described
 by
 a nearest neighbor tight-binding band.
We examined
 what kind of symmetry of the superconducting state
 could appear
 when the chemical potential $\bar{\mu}$
 and the mixing parameter $s$ of the $s$-channel interaction
 (equivalently, the mixing parameter $d=1-s$
 of the $d$-channel interaction)
 are varied.
The result
 was summarized
 into a phase diagram
 for the pairing symmetry
 in the $\bar{\mu}-s$ parameter plane.
The boundary line
 between the $s$-wave and the $d$-wave pairing region
 in the $\bar{\mu}-s$ parameter plane
 has been
 determined
 except for
 the
 region
 where $\bar{\mu}$ is less than 0.5
 and $s(d)$ is very small(large).
It seems
 that there is a lower bound in $\bar{\mu}$
 for the occurrence
 of the $d$-wave superconductivity.
Because of the limitation of the computational time,
 we could not give
 a decisive answer
 to
 the
 question
 whether there is
 the possibility
 for the occurrence
 of the $d$-wave superconductivity
 when $\bar{\mu}$ is small
 and $d(=1-s)$ is very large.
But we are convinced,
 from several trial calculations
 at very low temperature,
 that $T_{c}$ would be extremely low
 if the $d$-wave superconductivity
 could ever occur
 for $(\bar{\mu},s)$ in such a region.
Therefore,
 we can almost say
 that there is a threshold value of $\bar{\mu}$
 below which
 the $d$-wave superconductivity
 hardly
 occurs
 even when the $d$-channel interaction
 is
 mixed strongly.
On the other hand,
 the $s$-wave superconductivity
 can occur
 for any value of $\bar{\mu}$:
Even for large $\bar{\mu}$ near $\bar{\mu}=2$(half-filling case),
 it can occur
 if the $s$-channel interaction is sufficiently strong
 compared with the $d$-channel one.

Although the area of the $d$-wave pairing
 in the $\bar{\mu}-s$ plane
 is relatively small
 and spread only around the corner $(\bar{\mu},s)=(2,0)$,
 once the $d$-wave superconductivity occurs,
 it holds
 the advantage of getting a high $T_{c}$
 over the $s$-wave superconductivity.
In the close vicinity of
 half-filling $(\bar{\mu}=2)$,
 the effective coupling strength of the $d$-channel interaction
 is strongly enhanced
 and $T_{c}$ goes up rapidly
 as $d$ increases with a fixed $\bar{\mu}$
 and
 it
 does
 the same
 as $\bar{\mu}$ increases
 with a fixed $d(=1-s)$.
Within our model calculation,
 therefore,
 the case of a pure $d$-channel interaction
 and
 half-band filling
 is the most advantageous one
 so as to achieve a high $T_{c}$ superconductivity.


\vspace{10mm}
\noindent
\begin{Large}
{\bf Acknowledgements}
\end{Large}

This work was partially supported
 by the project for Parallel Processing
 and Super Computing at Computer Centaur University of Tokyo.
 The authors would like to thank Dr. Y. Shiina
 for helpful advices.

\newpage


\newpage


\setcounter{page}{1}
\renewcommand{\thepage}{\roman{page}}

\noindent
\begin{Large}
{\bf Figure captions} \\
\end{Large}

\begin{itemize}

\item[ Fig.1.]
Model function of $\overline{\alpha^2F(\Omega)}$.

\vspace{10mm}

\item[ Fig.2.]
Diagram for the specification of states (a)at 20[K] and (b)at 40[K].
Circles and triangles represent points $(s,\bar{\mu})$
 for which the states are superconducting
 with the $s$-wave pairing symmetry and with the $d$-wave one, respectively,
 and squares indicate points
 for which the states are normal at 20[K] in (a) and at 40[K] in (b).

\vspace{10mm}

\item[ Fig.3.]
Phase diagram of the pairing symmetry
 in the $\bar{\mu}-s$ plane.
The region hatched by vertical dotted lines
 is that of the $s$-wave symmetry
 where sets of the coordinates $(\bar{\mu},s)$
 can produce $s$-wave superconductivities.
The region hatched by horizontal dotted lines
 is that of the $d$-wave symmetry.
A thick solid line
 represents
 the boundary between regions
 of the $s$-wave symmetry and of the $d$-wave one.
Several thin solid lines drawn together are equi-$T_{c}$ lines,
 on each side
 of which the value of $T_{c}$ is indicated.

\vspace{10mm}

\item[ Fig.4.]
Curves of $T_{c}$ versus $s$
 for fixed $\bar{\mu}$ values,
 1.5, 0.8 and 0.5.
In each curve,
 the part drawn in solid line
 exhibits $T_{c}$ values
 of the $s$-wave superconductivity
 and the part
 in dashed line
 does $T_{c}$ values of the $d$-wave superconductivity.

\end{itemize}

\end{document}